\documentclass[conference]{IEEEtran}
\usepackage[hyphens]{url}
\usepackage{hyperref}
\hypersetup{colorlinks,allcolors=black}
\usepackage[backend=biber, style=ieee]{biblatex}

\usepackage{amsmath,amssymb,amsfonts}
\usepackage{algorithmic}
\usepackage{graphicx}
\usepackage{textcomp}
\usepackage{xcolor}
\usepackage{balance}
\usepackage{booktabs}
\usepackage{tikz}
\usepackage{array}
\usepackage{float}
\usepackage{geometry}
\begin{document}

\title{Transforming Information Systems Management: A Reference Model for Digital Engineering Integration}

\author{\IEEEauthorblockN{John Bonar}
	\IEEEauthorblockA{
 \textit{The Beacom College of Computer and Cyber Sciences}\\
 \textit{Dakota State University}\\
		Madison, SD, United States of America \\
		john.bonar@trojans.dsu.edu}\\
	\and
	\IEEEauthorblockN{John Hastings}
	\IEEEauthorblockA{
 \textit{The Beacom College of Computer and Cyber Sciences}\\
 \textit{Dakota State University}\\
		Madison, SD, United States of America \\
		john.hastings@dsu.edu}
}
\maketitle

\begin{abstract}
Digital engineering practices offer significant yet underutilized potential for improving information assurance and system lifecycle management. This paper examines how capabilities like model-based engineering, digital threads, and integrated product lifecycles can address gaps in prevailing frameworks. A reference model demonstrates applying digital engineering techniques to a reference information system, exhibiting enhanced traceability, risk visibility, accuracy, and integration. The model links strategic needs to requirements and architecture while reusing authoritative elements across views. Analysis of the model shows digital engineering closes gaps in compliance, monitoring, change management, and risk assessment. Findings indicate purposeful digital engineering adoption could transform cybersecurity, operations, service delivery, and system governance through comprehensive digital system representations. This research provides a foundation for maturing application of digital engineering for information systems as organizations modernize infrastructure and pursue digital transformation.
\end{abstract}


\section{Introduction}

As organizations increasingly depend on secure and resilient information systems that evolve with business needs, the challenges of traditional document-centric approaches become apparent. These methods often fall short in providing the necessary visibility, adaptability, and integration required for effective system design, security, and lifecycle management. Digital engineering emerges as a transformative discipline designed to address these shortcomings through the comprehensive digital representation of systems. This approach facilitates enhanced visibility, integration, automation, and analytics across the entire lifecycle of information systems.

This paper explores the application of digital engineering to improve information assurance and system lifecycle management. By leveraging a model-based approach, which incorporates the core principles of model-based systems engineering
, Digital Threads (DTh), and Digital Twins (DTw), the methodology extends beyond conventional capabilities to offer a robust framework for managing complex systems more efficiently.

By integrating these digital engineering practices within existing information assurance frameworks and system lifecycle methodologies, we aim to significantly enhance the security, compliance, risk management, service delivery, and continuity of operations of information systems. The subsequent sections will delve into how these integrated practices can close gaps in current frameworks, backed by a reference model that demonstrates their application in a real-world context. 

\section{Background}\label{background}

This section details background related to digital engineering, information assurance,
and information system lifecycle management, providing a 
foundation for the reference model and analyses
presented later in this paper.

\subsection{Digital Engineering}

Digital engineering (DE) represents a paradigm shift in system design, development, sustainment, and lifecycle management. This discipline utilizes model-based approaches and digital technologies to transcend traditional document-centric methods \cite{Hutchison_Blackburn_Clifford_Yu_Chen_Tech_Salado_Henderson_McDermott_Van_Aken_2020}, enabling enhanced visibility, integration, automation, and analytics. The core components of DE include:

\subsubsection{Model-Based System Engineering}

\citeauthor*{Bone_Blackburn_Rhodes_Cohen_Guerrero_2019} \cite{Bone_Blackburn_Rhodes_Cohen_Guerrero_2019} describes how model-based systems engineering (MBSE) captures system models rather than paper documentation across requirements \cite{ISO_15288_2023}, architecture \cite{Mahendra_Gaol_Supangkat_Ranti_2023,Masuda_Zimmermann_Shepard_Schmidt_Shirasaka_2021,Möhring_Keller_Schmidt_Sandkuhl_Zimmermann_2023}, design, integration, and testing. This allows analytical insight into system behaviors and provides a digital source of truth. MBSE enables integration with other systems, simulation, automated analysis, and requirements traceability. 

MBSE enables organizations to utilize object-based architectural frameworks that are most applicable to the organization such as:

\begin{itemize}
    \item Zachmann Framework \cite{Fatolahi_Shams_2006}
    \item Universal Architecture Framework (UAF) \cite{OMG_UAF_2022}
    \item UAF Enterprise Architecture (UAFEA) \cite{OMG_UAF_2022}
    \item Department of Defense Architecture Framework (DoDAF) \cite{DODA}
    \item The Open Group Architecture Framework (TOGAF) \cite{Josey_2023}
    \item North Atlantic Treaty Organization (NATO) Architecture Framework (NAF) \cite{NATO_AF_2022}
\end{itemize}

\subsubsection{Digital Threads}

DTh's connect digital artifacts across tools, stakeholders, and lifecycle stages. They provide authoritative sources of truth through traceability of relationships between elements like requirements, tests, issues, and code changes. DTh's are fundamental to managing complexity. \citeauthor*{Baker_Pepe_Hutchison_Tao_Peak_Blackburn_Khan_Whitcomb_2021} \cite{Baker_Pepe_Hutchison_Tao_Peak_Blackburn_Khan_Whitcomb_2021} explores the vital importance of DTh, which is further reinforced by the \citeauthor*{Engineer} \cite{Engineer} and the \citeauthor*{DODA} \cite{DODA} in how authoritative traceability is key to ensuring end-to-end management of any complex system.

\subsubsection{Digital Twins}

DTw's are virtual replicas of systems that can be integrated with physical instances. Twins allow low-risk testing \cite{Holmes_Papathanasaki_Maglaras_Ferrag_Nepal_Janicke_2021,Garrett_Kassan_2021}, prediction \cite{Engineer,DODA}, and evaluating changes \cite{Karaarslan_Babiker_2021, Garrett_Kassan_2021} before deployment. They support integration across domains like design, simulation, IoT, and analytics \cite{Khan_Saad_Niyato_Han_Hong_2022}. \citeauthor*{DeLorme_Clifford_2016} \cite{DeLorme_Clifford_2016} describes the value of DTw's in complex system of systems acquisition, and while their focus was on defense, \citeauthor*{DeLaurentis_Domercant_Guariniello_McDermott_Witus_2020} \cite{DeLaurentis_Domercant_Guariniello_McDermott_Witus_2020} describes similar benefits in a more general approach.

\subsubsection{Product Lifecycle Management}

Product lifecycle management (PLM) centers on managing information systems across concept, development, production, utilization, support, and retirement stages. PLM enables configuration control, change impact analysis, knowledge retention, and integration between disciplines. PLM is a critical function of modern cybersecurity and IT operations/service delivery frameworks such as the NIST Risk Management Framework (RMF) \cite{Force_2018,Force_2020}, and ITIL \cite{Cannon_AXELOS} as it would apply towards information systems.

Together, these capabilities aim to transform system engineering through comprehensive digital representations. DE unlocks new potentials in system governance \cite{Heihoff-Schwede_Kaiser_Dumitrescu_2019, Holmes_Papathanasaki_Maglaras_Ferrag_Nepal_Janicke_2021}, insight, automation \cite{Holmes_Papathanasaki_Maglaras_Ferrag_Nepal_Janicke_2021}, and integration. It requires changes in tools, policy, culture, and workforce skills \cite{Baker_Pepe_Hutchison_Tao_Peak_Blackburn_Khan_Whitcomb_2021} in order to realize full advantages. Adoption is accelerating in aerospace, defense, auto, and other industries to improve quality, costs, and delivery speed.

\subsection{Information Assurance}

Information Assurance (IA) involves practices that protect and defend information and information systems critical to an organization's mission. This includes ensuring confidentiality, integrity, authentication, availability, compliance, and non-repudiation. IA combines elements of cybersecurity, operational security, and risk management.

Key frameworks provide structured methodologies for implementing IA controls and processes. The NIST Cybersecurity Framework \cite{Cybersecurity_Framework_2013} offers standards and best practices for critical infrastructure sectors. It provides a risk-based taxonomy for assessing and improving cybersecurity programs. ISO 27001 \cite{ISO_27001} is an international standard focused on implementing information security controls based on organizational risk. It provides requirements for a comprehensive information security management system. The NIST RMF tailored for U.S. federal agencies, certain critical infrastructure, and the Defense Industrial Base. The RMF emphasizes near real-time or continuous risk assessment, authorization, and monitoring.

Underlying these frameworks and controls, risk management entails a continuous process of assessing threats and vulnerabilities to prioritize safeguards. Quantitative methods weigh likelihood and impact. Qualitative approaches leverage Delphi techniques tapping experts. Risk management directly feeds cybersecurity investment decisions and control selections.

Mature IA requires integrating frameworks, controls, and risk management into a holistic program. This program should embed into overarching enterprise risk governance and utilize automation \cite{Zimmerman_2019,Pandey_Wazid_Mishra_Mohd_Singh_2023} to enable continuous, scalable oversight. As threats evolve, IA must shift left earlier into system lifecycles and expand scope to new attack surfaces.

\subsection{Information System Lifecycle Management}

Information system lifecycle management (ISLM), or Information Technology Lifecycle Management, involves structured approaches for governing systems across concept, development, implementation, operation, maintenance and retirement. ISLM aims to optimize information technology service delivery while managing risk.

The Information Technology Infrastructure Library (ITIL) framework \cite{Cannon_AXELOS} offers a popular ISLM methodology used globally. ITIL covers the full service lifecycle \cite{Cannon_AXELOS} including strategy, design, transition, operations, and continual improvement. It provides best practices that organizations can tailor and adopt incrementally. Benefits include improved quality, cost optimization, risk reduction, and business alignment.

ISLM includes interrelated functions like change management, release management, service portfolio management, and knowledge management. Change management \cite{Farenden_2012} assesses impact and risk to minimize disruption from system changes. Release management deploys new functionality and fixes in a controlled manner. Portfolio management aligns IT investments with business priorities. Knowledge management retains information across staff turnover.

Strong ISLM integrates with information security programs \cite{ISO_27001,Force_2018,Force_2020,Cannon_AXELOS} for risk-based oversight. It leverages configuration management databases for near real-time system visualization. Agile systems development extends ISLM techniques for faster delivery with security and reliability. As environments grow more complex, automation and analytics become critical ISLM capabilities.

\begin{figure}
    \centering
    \includegraphics[width=1\columnwidth, keepaspectratio]{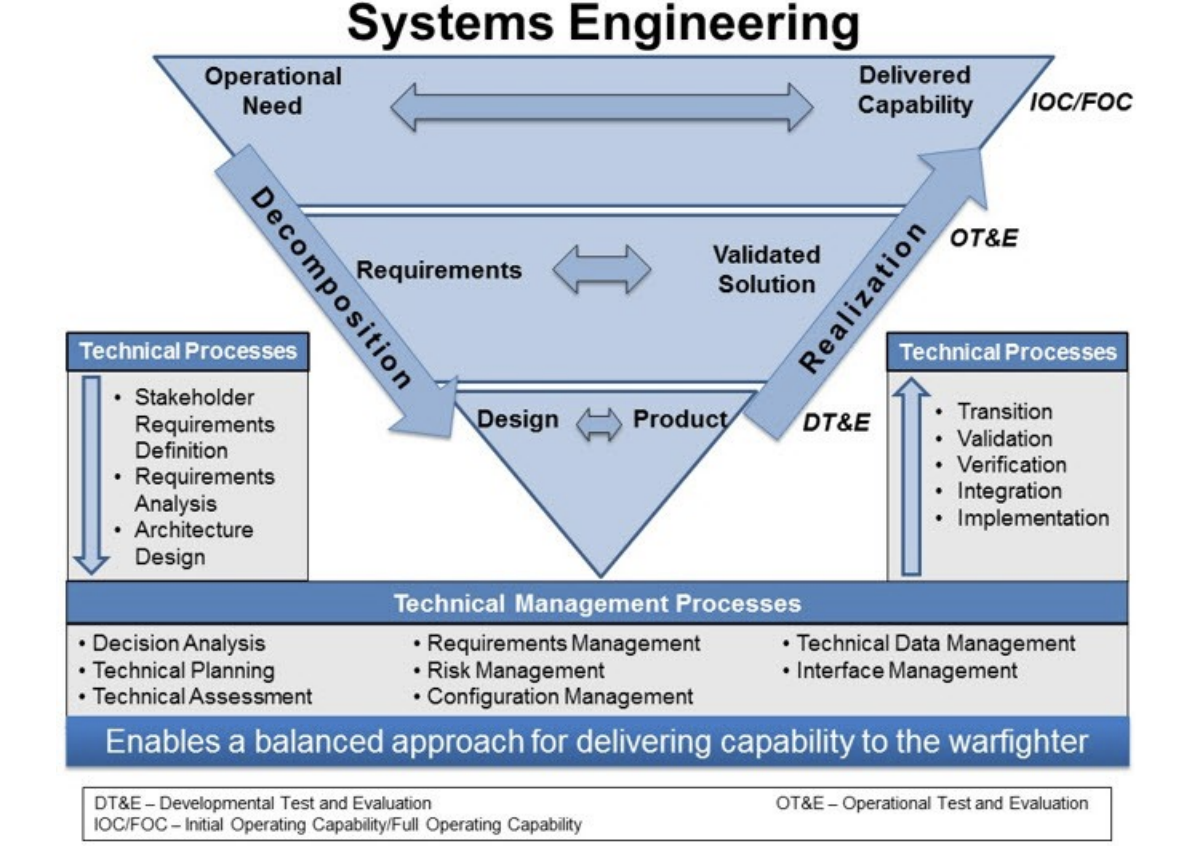}
    \caption{Department of Defense Systems Engineering Lifecycle \cite{DoD_Research_and_Engineering_2022}}
    \label{fig:engineeringv}
\end{figure}

Information systems lifecycle management frameworks such as ITIL \cite{Cannon_AXELOS} have several steps that align with the NIST RMF, and the Department of Defense Systems Engineering Lifecycle \cite{DODA}. This alignment of lifecycles appears to show a synergy with the nature of systems engineering disciplines that allow for a product or platform agnostic approach. Figure \ref{fig:engineeringv} 
visualizes the Department of Defense \cite{DODA} systems engineering lifecycle which aligns to other ISLM life cycles in a broad comparison.

While approaches continue maturing, ISLM forms a key pillar of IT service delivery. It enables managing technology assets across their entire lifecycle. ISLM integration with frameworks such as ITIL provides a strategic advantage in controlling costs while enabling innovation. \citeauthor*{Jamous_Bosse_2016} \cite{Jamous_Bosse_2016} explores how various frameworks interrelate across silos. 

\begin{figure}
    \centering
    \includegraphics[width=1\columnwidth, keepaspectratio]{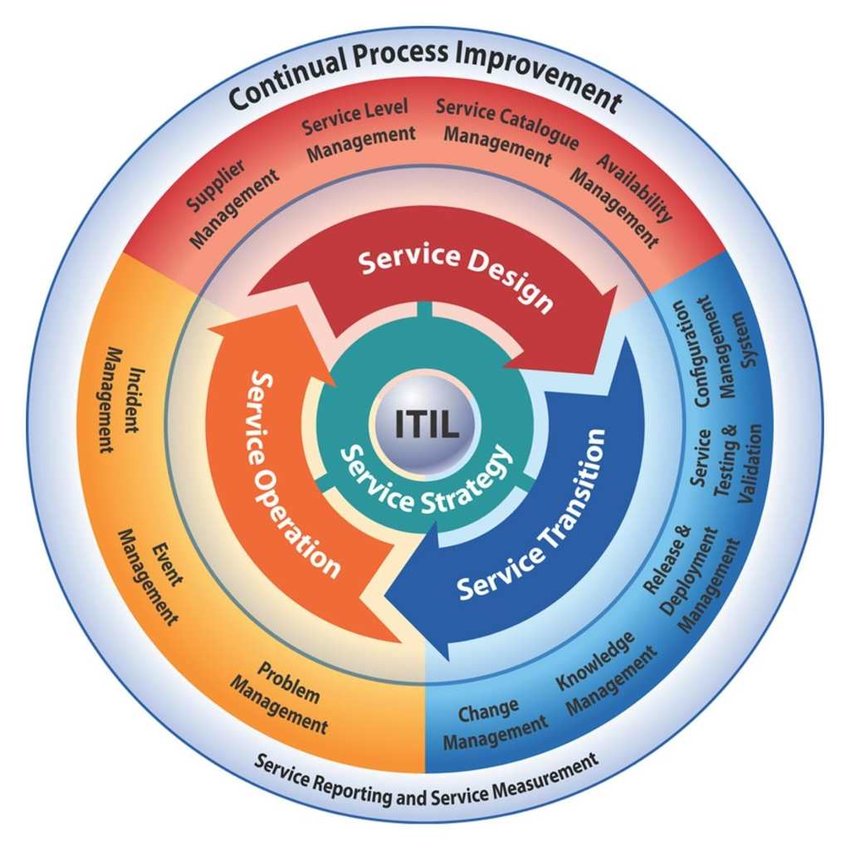}
    \caption{ITIL Lifecycle \cite{Cannon_AXELOS}}
    \label{fig:ITIL_Lifecycle}
\end{figure}

\begin{figure}
    \centering
    \includegraphics[width=1\columnwidth, keepaspectratio]{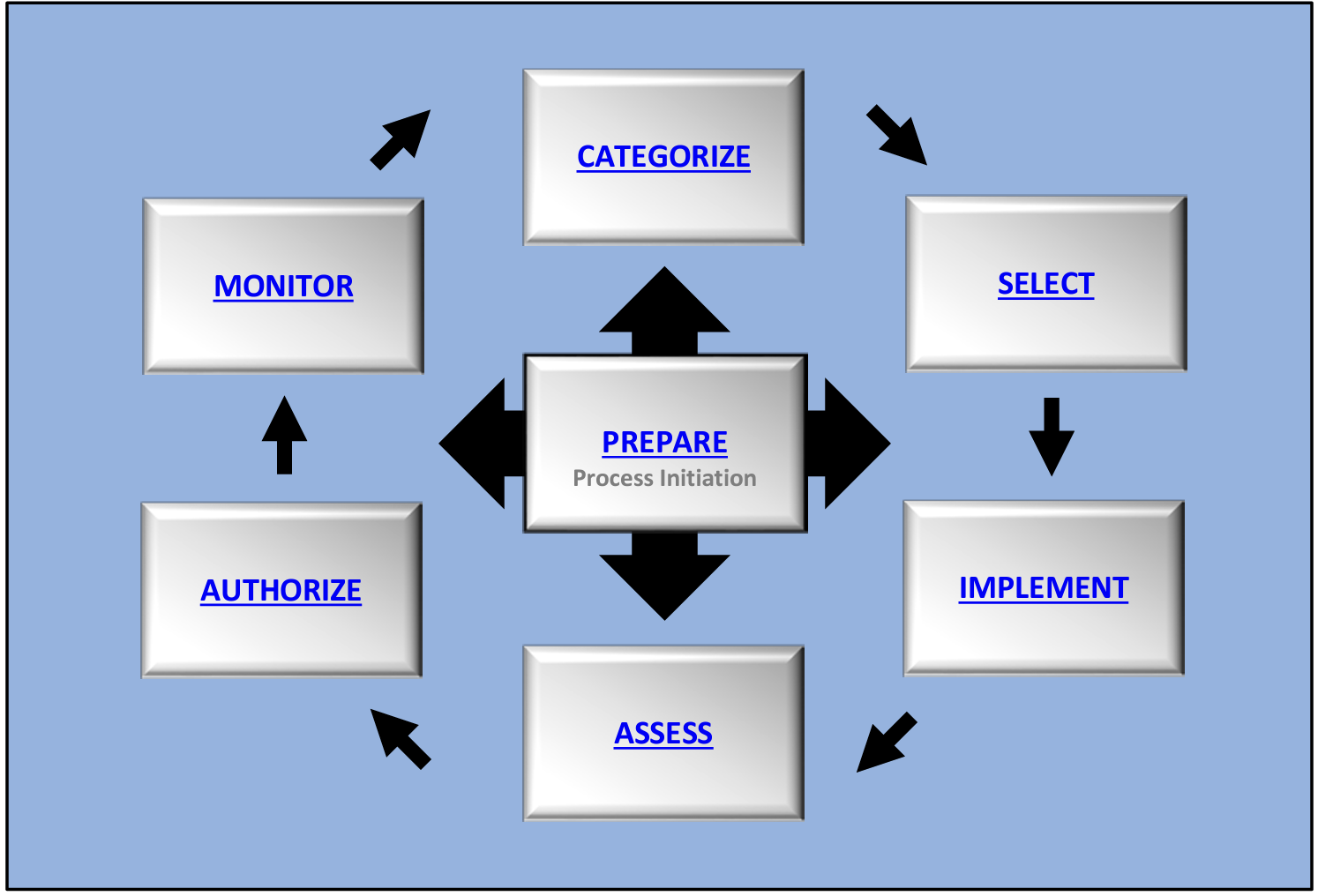}
    \caption{RMF Lifecycle \cite{Force_2018}}
    \label{fig:RMF_Lifecycle}
\end{figure}

The continuous lifecycles of the DoD System Engineering (Fig. \ref{fig:engineeringv}), ITIL (Fig. \ref{fig:ITIL_Lifecycle}), and RMF (Fig. \ref{fig:RMF_Lifecycle}) frameworks follow similar steps at a high level. Each framework is focused on specific areas such as Systems Engineering, Service Delivery, and IA. The value of specialized frameworks enables value for organizations, but if they are executed in isolated silos then organizations will continue to realize inefficiencies and inability to visualize operational risks.

\section{Background Analysis}\label{background analysis}

A review of each pillar of DE clearly reveals that they are interconnected via dependencies and cross-functional requirements. These derived capabilities represent requirements that derive from a particular capability, for example, MBSE is a derived core requirement to realize a DTh capability. This requirement on MBSE by DTh is clearly visible as DThs rely on object-based artifacts to enable the traceability capability that is at the heart of DTh.\footnote{ Derived requirements are omitted from the analysis findings to ensure a clear distinction between what existing IA and ISLM frameworks require as a capability, versus what is needed to realize that capability.}

An analysis of IA and ISLM frameworks (from the prior section) reveals that only the RMF and ITIL have requirements for some aspect of DE capabilities. In contrast, DE capabilities can be aligned to any of the IA and ISLM frameworks selected in this research. Table \ref{tab:Framework-Comparison} summarizes the requirements identified within each framework as related to DE capabilities. 
The PLM capability is required to fully meet certain controls within all IA and ISLM frameworks, with the NIS RMF requiring DTh capability as part of the frameworks continuous monitoring language.

\begin{table}[H]
        \centering
	\caption{Digital Engineering Requirements Comparison }
	\begin{tabular}{@{}ccccc@{}}
	\toprule
	\textbf{Framework} & \textbf{MBSE Req} & \textbf{DTh Req} & \textbf{DTw Req} & \textbf{PLM Req} \\ \midrule
	\textbf{DoD DE}    & Yes & Yes & Yes & Yes \\
	\textbf{NIST CSF}  & No  & No  & No  & Yes  \\
	\textbf{ISO 27001} & No  & No  & No  & Yes  \\
	\textbf{NIST RMF}  & Yes  & Yes  & No  & Yes  \\
	\textbf{ITIL}      & No  & No  & No  & Yes  \\ \bottomrule
	\end{tabular}
	\label{tab:Framework-Comparison}
	\end{table}

According to our analysis, there are a number of gaps within 
the current IA and ISLM frameworks that DE can address. Figure \ref{ia and islm gaps} summarizes the identified gaps specific to each pillar of DE along with corresponding recommendations.

It is important to note that, of the frameworks investigated in this research project, some aspects of DE capabilities are evident to some degree. However, the competencies and capabilities of many organizations currently are not able to effectively implement such capabilities for a number of reasons whether it be funding, awareness, or lack of perceived value by the organization. Regardless of the reason, there are organizations that will adopt DE in support of IA and ISLM improvement, and improved visibility for their respective enterprises.

\begin{figure}
  \scriptsize
\begin{enumerate}
	\item Model-Based System Engineering
    \begin{enumerate}
	\item Gap: Existing frameworks often rely on static documentation that does not support real-time updates or automated processes, hindering proactive risk management and compliance monitoring.
        \item  Recommendation: Adopt MBSE to ensure dynamic system modeling which enhances traceability, supports automated testing and analysis, and integrates with existing frameworks to improve real-time decision-making and compliance.
    \end{enumerate}
	\item Digital Threads
    \begin{enumerate}
        \item Gap: The absence of DTh leads to fragmented system visibility and manual traceability efforts, increasing the risk of errors and oversight in complex system interactions.
        \item Recommendation: Implement DTh to establish a continuous link across different lifecycle stages, enhancing the traceability of requirements, designs, and operational metrics. This ensures cohesive system oversight and facilitates impact analysis of changes.
    \end{enumerate}
	\item Digital Twin 
    \begin{enumerate}
        \item Gap: Current practices do not fully leverage simulation for preemptive testing and evaluation, reducing the system's resilience to adapt to changes and anticipate future challenges.
        \item Recommendation: Utilize DTw to mirror physical and digital systems, allowing for safe testing of potential changes in a controlled virtual environment, thereby reducing downtime and enhancing system preparedness for real-world scenarios.
    \end{enumerate}
	\item PLM
    \begin{enumerate}
        \item Gap: Inadequate integration of lifecycle stages with compliance and risk management processes often results in inefficiencies and lack of responsiveness to emerging threats or changes.
        \item Recommendation: Enhance PLM practices to tightly integrate with digital engineering tools like DTw and DTh, improving the visibility and management of changes throughout the product lifecycle, thereby increasing agility and compliance.
        \end{enumerate}
\end{enumerate}
\caption{IA and ISLM Gaps \& Recommendations Within Each Pillar of DE}
\label{ia and islm gaps}
\end{figure}

\section{Reference Model}

Based on the analysis of the current landscape, a design science methodology was employed to produce the reference model shown in Figure \ref{reference model}. The reference model utilizes MBSE tools that demonstrate object-based elements from concept, requirements, architecture, design, and implementation. This model focuses on establishing traceability between elements, and views.\footnote{Sustainment, retirement, disposal are not included in the model at this time, and are addressed in the future work section.} 
The reference model focuses on Client and Application Network Connectivity for Identity and notional ISB application, a selection of Security Controls of the client, switch, and servers for Application and Authentication. The ISB is a notional organization with an application utilized by staff to conduct work on widgets.

\begin{figure}
  \scriptsize
\begin{itemize}
	\item 1 Windows Workstation
	\begin{itemize}
        \item Running a notional ISB Client Application
    \end{itemize}
	\item 1 Network Switch
	\item 1 Windows Server Application Server
    \begin{itemize}
    	\item Notional ISB Server Application
    	\item PLM Application
        \item DTw Service
        \item DTh Service
        \item MBSE Server Application
    \end{itemize}
	\item 1 Windows Server Domain Controller
    \begin{itemize}
        \item Active Directory Domain Services
        \item Authentication Services
    \end{itemize}
    \item 1 Windows Server Security Server
    \begin{itemize}
        \item SEIM Application
        \item Vulnerability Scanning and Management Application
        \item Monitoring Application
    \end{itemize}
\end{itemize}
\caption{Reference Model Components}
\label{reference model}
\end{figure}

\subsection{Conceptual/Strategic and Requirements Management}

Conceptual needs are captured as unique elements within the model itself as Strategic, Business, and Security elements. These elements are utilized to create requirements with traceable associations that can be ``reviewed'' to ensure that the concepts are satisfied by notional leadership, customers, or auditors. Due to limitations of budget and scope, a formal requirements management solution such as Doors Next Gen will not be utilized and will be captured and managed within the MBSE model.

\subsection{Architecture and Design}

Architecture and Design views focus on various layers specific to the Information System, and objects are reused across multiple views to ensure traceability within the model. Object reuse ensures that elements maintain relationships, are authoritative, and traceable. By utilizing authoritative objects, this will ensure as the Information System is deployed configurations may be utilized in a repeatable and scalable manner to ensure that as the organization evolves the model can act as the authoritative source of truth for IA and ISLM.

\subsection{Implementation and Sustainment}
Implementation and Sustainment Views focus on logical flows of applications to demonstrate how the Information System can realize DE capabilities within the Information System in support of IA and ISLM.

\section{Reference Model Views}
This section presents an illustrative overview of the reference model in the form of a variety of views. Due to the complexity of a model when considering the many aspects of even a small information system, this overview focuses on the select views and context
rather than deeper dive.

\subsection{Conceptual/Strategic View}

\begin{figure}
    \centering
    \includegraphics[width=1\columnwidth, keepaspectratio]{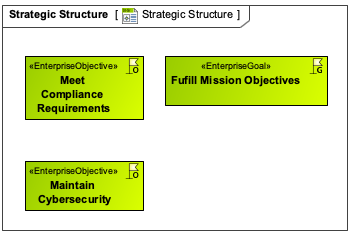}
    \caption{Conceptual/Strategic View}
    \label{fig:Conceptual View}
\end{figure}

The conceptual/strategic view articulates the conceptual desires of the organization. The reference model, presented in Figure \ref{fig:Conceptual View}, encapsulates the ISB's enterprise goal to fulfill mission objects, and two enterprise objectives: maintain cybersecurity and meet compliance requirements. This strategic framework sets the stage for translating broad organizational aims into specific, actionable requirements. It serves as a critical reference point for aligning technological capabilities with business objectives, ensuring that each component of the system contributes effectively to the enterprise's overarching goals. In practical terms, this conceptual layer incorporates multiple strategic elements which, although numerous in a real-world setting, are distilled here to exemplify the most critical aspects.

\subsection{Requirements View}

\begin{figure}
    \centering
    \includegraphics[width=1\columnwidth, keepaspectratio]{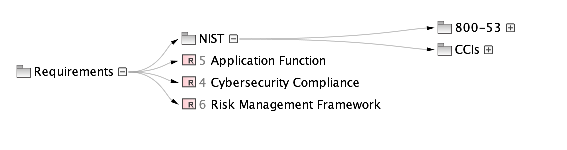}
    \caption{Top-Level Requirements View}
    \label{fig:Top-Level Requirements View}
\end{figure}

\begin{figure}
    \centering
    \includegraphics[width=1\columnwidth, keepaspectratio]{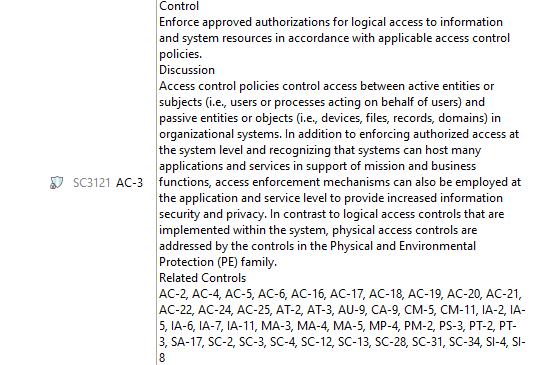}
    \caption{Requirements Table View}
    \label{fig:Requirements Table View}
\end{figure}

\begin{figure}
    \centering
    \includegraphics[width=1\columnwidth, keepaspectratio]{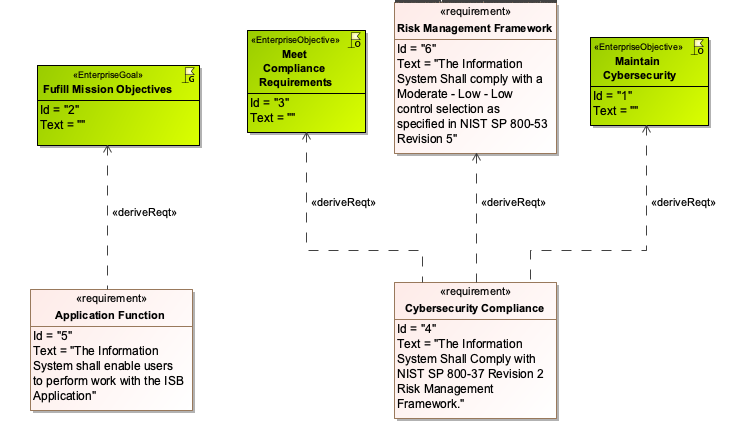}
    \caption{Requirements Table View}
    \label{fig:Requirements Traceability View}
\end{figure}

The requirements view delves into the complexity of translating strategic intents into specific, actionable requirements. Within this level, the number of views and elements begins to grow by several orders of magnitude. In this reference model, the ISB Organization employs a NIST RMF Based Cybersecurity and Compliance approach, which in turn supports defining certain top-level requirements (shown in Fig. \ref{fig:Top-Level Requirements View}), such as application functionality of the ISB application, and RMF. From this, the RMF requirement allows the model to draw requirements from NIST SP 800-53 Revision 5. As seen in Figures \ref{fig:Requirements Table View} and \ref{fig:Requirements Traceability View}. Given the number of controls that are being treated in this model as Cybersecurity Requirement types, and the Common Correlation Identifiers (CCIs) as derived Cybersecurity Requirement types, the number of requirement element within the model swells to over three thousand elements within the requirements container of the model. The inclusion of relationships between the requirements including indirect relationships brings the number of elements to over 18,000.

Due to the large number of elements complete views of all requirements elements is not possible in this paper, instead Figure \ref{fig:Requirements Traceability View} represents the top two levels of requirements of the ISB organization with tier three requirements from the RMF element which serves as the outgoing ``Derived From'' relationship to each NIST SP 800-53 Revision 5 Controls and Control Enhancements.

\subsection{Application Architecture Views}

\begin{figure}
    \centering
    \includegraphics[width=1\columnwidth, keepaspectratio]{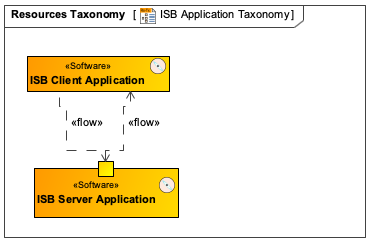}
    \caption{ISB Application View}
    \label{fig:ISB Application View}
\end{figure}

The application view focuses on various application-level aspects, such as application interactions, authentication, and logical topology of the Information System. The interaction of the ISB application is illustrated in Figure \ref{fig:ISB Application View}. This view contains the elements and relationships of the elements, however within these elements various data points and information may be captured such as: Ports, Protocols, Information Exchanges, Documentation, and Relationships to other elements within the model. This enables staff to reuse elements within the model and only display the specific aspects desired for the context within a view.

\begin{figure}
    \centering
    \includegraphics[width=1\columnwidth, keepaspectratio]{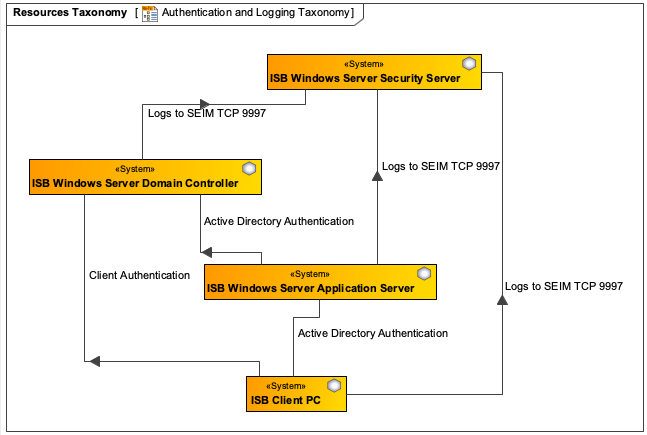}
    \caption{Authentication and Logging Application View}
    \label{app-auth-log}
\end{figure}

A more detailed application view of the Authentication and Logging View is shown in Figure \ref{app-auth-log}. Within this view we see a complete representation of the server and client infrastructure.\footnote{The network switch is missing in this view as it only serves as a transport fabric and is not necessary for this view.}

\begin{figure}
    \centering
    \includegraphics[width=1\columnwidth, keepaspectratio]{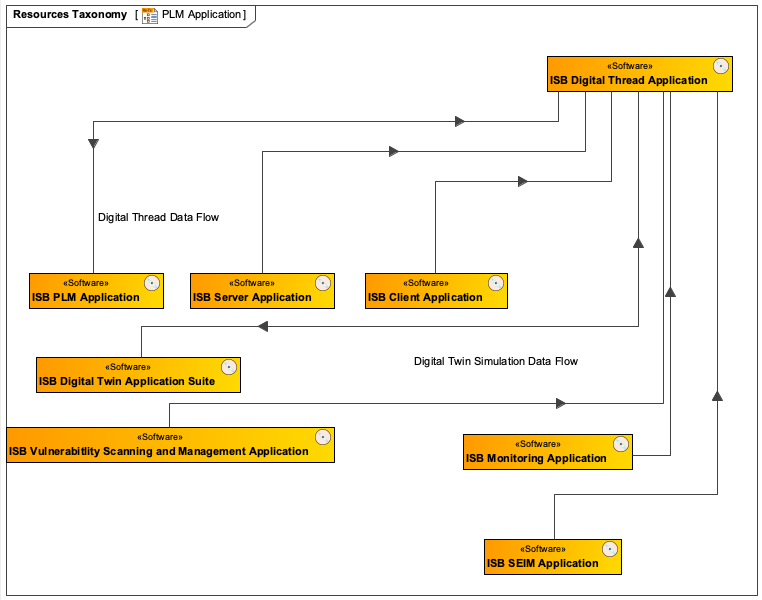}
    \caption{Digital Engineering Pillars Application View}
    \label{de-app-view}
\end{figure}

Figure \ref{de-app-view} illustrates a higher-level view in which the data interactions of PLM integrate into the ISB Information System in along with a DTh Solution and DTw solution to realize DE capabilities. This view provides a logical flow of data between the various applications within the Information System of the ISB organization. It is in this view that we take notice that the DTh solution is the central component, it is this application itself that acts as a translator and data flow manager to ensure that different applications can to establish traceability between different DE pillars and applications that enable those capabilities.

In order to properly realize DE within the notional ISB organization, several non-DE systems are utilized such as Monitoring, SEIM, and Vulnerability and Scanning Management. These applications have existing APIs that the DTh solution can translate and flow to the PLM suite and DTw solution. This enables the ISB organization to interlink multiple sources of truth and to realize a highly robust capability to demonstrate with high assurance the state of their Information System in terms of cybersecurity, compliance, and operations.

\subsection{Technical Architecture \& Detailed Design View}

\begin{figure}
    \centering
    \includegraphics[width=1\columnwidth, keepaspectratio]{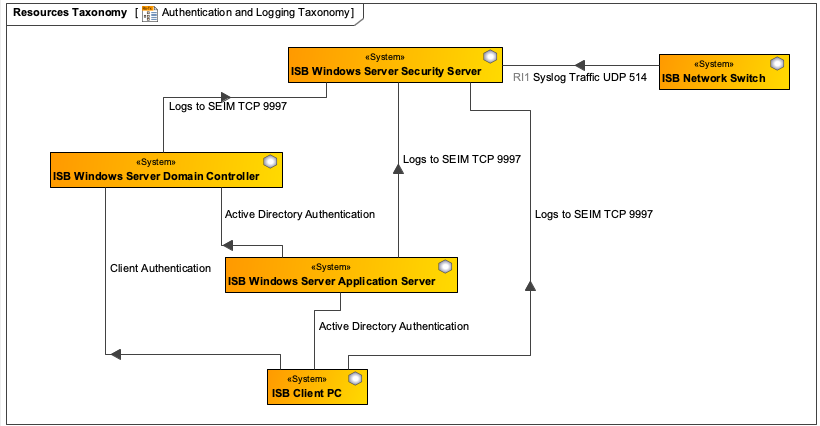}
    \caption{Authentication and Logging Technology View}
    \label{tech-auth-log}
\end{figure}

This view captures the details of the technical architecture and design elements that underpin the information system.
The view offers organizations the ability to utilize multiple views with element reuse, allowing for multiple stakeholders within the organization to view specific aspects of the information system that display relevant information appropriate to the view. Figure \ref{tech-auth-log} serves as an example to demonstrate the Authentication and Logging process at a technology level. This view provides additional technology-related data compared to a similar view in the application layer.

\begin{figure}
    \centering
    \includegraphics[width=1\columnwidth, keepaspectratio]{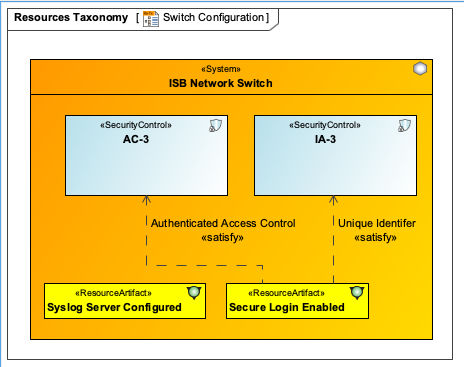}
    \caption{Network Switch Configuration Technology View}    
	\label{fig:Network Switch Configuration Technology View}
\end{figure}

With each element able to contain data, sub elements, relationships, tags, documentation, and other information fields, this capability allows for elements to serve as artifacts for an Information System Security Plan Body of Evidence. Figure \ref{fig:Network Switch Configuration Technology View} depicts how a sub element within the network switch element represents a configuration that satisfies two cybersecurity requirements.

\section{Model Evaluation}\label{evaluation}

This section describes an evaluation of the reference model to ensure that it properly addresses the gaps identified in Section \ref{background analysis}. Such evaluation plays a pivotal role in identifying potential areas for improvement, validating alignment with practices \& standards, and assessing the model's ability to serve as a reliable template for future implementations.

\subsection{Addressing Gaps}

A continual challenge with current assurance and lifecycle practices involves limited end-to-end visibility across layers \cite{Masuda_Zimmermann_Shepard_Schmidt_Shirasaka_2021,DeLorme_Clifford_2016,Shao_2021,Rhodes_2020}, from strategic drivers down to configured devices. Documents abstracted from implementations disconnect as systems continuously evolve. The reference model introduced in this research addresses this reality through integrated DTh's linking strategic goals to satisfied technical requirements and deployed configurations. This traceability enables downstream impact analysis when contextual changes occur.

Siloed teams, processes, and tools represent another gap \cite{Hutchison_Verma_Burke_Tao_Giffin_Yu_Makwana_Chen_Xiao_2020} unable to provide a holistic perspective. The model connects disparate data sources across domains \cite{DeLaurentis_Domercant_Guariniello_McDermott_Witus_2020} whether requirements management, application lifecycles, or infrastructure topology. Composing authoritative system representations breaks down barriers between disciplines. Component relationships reveal cross-domain interactions not visible within knowledge silos.

Manual verification and auditing of compliance controls presents an additional shortcoming. The model allows managing configurations through standardized objects mapped to requirements. Design patterns automate building secure architectures. Automated dependency analysis supplements human-driven reviews, enabling continuous assurance monitoring.

Overall, the model demonstrates how DE practices applied holistically help address limitations constraining current IA and lifecycle methodologies. Transitioning to accurate living models provides the visibility, integration, and automation needed to manage digital infrastructure complexity with security, resilience, and agility.

\subsection{Traceability and Object Reuse}

A key advantage exhibited by the model is its end-to-end traceability across layers and elements made possible through integrated DTh's. Strategic concerns connect downstream to system requirements. Requirements map to specific architecture components and configurations that implement them. This provides a level of visibility unachievable with traditional document-based approaches prone to configuration errors and drift. Changes to strategic drivers or requirements cascade effects across DTh relationships to assess impacts throughout the system lifecycle.

Furthermore, the model maintains authoritative sources of truth by reusing model elements across multiple views. An application server element, for example, represents the same entity across conceptual, logical, physical, and implemented layers. This avoids disconnected duplicates as architectural layers transform from abstract to concrete. Component relationships are unambiguous as single definitions propagate across views tailored to stakeholder perspectives. 

Each instance of the application server instance may have different configuration sub elements applied to the instance of the application server instance. This allows for the use of element stereotypes and common sub element configurations that are applied to all application server instances. This allows for organizations to ensure common baselines are applied, and with the use of DTh's the instantiation of each element instance may be verified as a DTw or actual implementation.

Standardized configurations assigned to elements also reinforce consistency. Servers adopt defined configuration elements that embed policies, baselines, and compliance requirements. As configurations change, updates apply across assigned elements rather than individually configured assets that drift. Model-based reuse avoids redundancy while enabling rapid change analysis not possible when managing individual documents or repositories.

In summary, integrated DTh's and authoritative object reuse exhibited by the model illustrate the significant benefits compared to traditional document-centric approaches. The model demonstrates traceability, accuracy, and integration essential for managing complexity at scale.

\subsection{Application of DE Practices}

The reference model provides a limited yet tangible example of core DE practices applied to enhance IA and lifecycle management. Model-based techniques capture system composition and configurations digitally for analytical insight versus traditional document-based approaches. Strategic goals connect downstream through DTh's into hardware and software elements that implement requirements traceably. This end-to-end visibility enables assessing the system holistically and understanding change impacts across layers.

DTw capabilities are reflected in the ability to evaluate alterations to the environment by propagating effects through component relationships prior to deployment. The model encapsulates the entire product lifecycle by connecting conceptual needs to sustainment and configurations. Integration of requirements management into design activities breaks down silos, while architecture views provide multidimensional visualizations across abstraction levels. Standardized configuration objects enforce consistency for servers, networks, and other asset types.

While limited in scope, the model indicates the art of the possible by demonstrating foundational DE methods. As complexity scales across integrated systems, the need for accurate digital representations will only increase. This research provides a foundation for maturing techniques that leverage the full advantages of model-based approaches for security, operations, and management.

\section{Future Work}\label{future}

Significant opportunities exist to mature the reference model into an operational DTw and advance adoption of DE in IA and lifecycle management. The following topics represent considerations of future work and research to mature this research topic.

\subsection{Continued Reference Model Development}

Continuing to develop and implement additional components like infrastructure, applications, requirements, views, automation, sustainment, retirement, and disposal will enable the reference model to serve as an exemplar for reducing the initial effort required to create models for organizations. By developing a reusable reference model as a template, teams and individuals can reduce errors when instantiating their own DE transformation initiatives.

\subsection{Digital Twin of Reference Model}

Developing a functional model with simulation capabilities could demonstrate significant opportunities for incident response, disaster planning, risk assessment, and operational optimization. By enabling organizations to simulate multiple scenarios in a synthetic model, teams could quickly understand where resilience, security incidents may create cascading failures, or how such disruptions could be avoided.

\subsection{Model Automation and Real-Time Integration}

Automating model generation for technical and application layers is another promising avenue. Enabling organizations to utilize an automated interface that can ingest and translate organizational infrastructure, partially or wholly, would benefit them by ensuring a real-time understanding of how the information system exists in its current temporal state. This vision of full end-to-end traceability and lifecycle management appears to offer significant benefits across multiple disciplines.

\subsection{Value and Viability Study}

Understanding how the application of DE capabilities is a transformational effort for organizations, requiring knowledge, tools, culture, and capabilities to implement and maintain. Investigating how DE can deliver value and the viability for organizations represents an undertaking that would justify further research into this approach. Given the complex nature of mid to large sized organization technology portfolios.

\section{Conclusion}\label{conclusion}
This paper demonstrates the significant benefits DE practices can offer for IA and lifecycle management. The reference model provides a practical example of applying DE techniques to an information system context.

Integrating DTh's, twins, MBSE, and PLM allows organizations to achieve end-to-end traceability, improved risk management, optimize ISLM, and enhanced operational visibility. The model exhibits these capabilities in a limited but concrete manner. DE closes gaps in connecting teams, data, and tools across security, operations, and management.

Transitioning from document-based to model-based approaches enables real-time accuracy, impact analysis, automation, and integration. Models provide single sources of truth and authoritative system representations. This research indicates DE could transform the ability to visualize, analyze, and optimize information systems if purposefully applied.

While progressing maturity requires investment and organizational change, the model and analysis illustrate DE’s immense potential. Adoption stands to significantly enhance cybersecurity, compliance, system quality, and lower lifecycle costs. As organizations modernize infrastructure and embrace digital transformation, leveraging DE techniques will become increasingly imperative.

This paper provides a foundation for future research to further refine use of models for managing system lifecycles. The possibilities of DE to fundamentally transform how we conceive, design, implement, and manage information systems are worthy of continued exploration. With collaboration between researchers and industry, this discipline’s full benefits can be progressively realized.

\section*{Acknowledgments}
Claude \cite{claude} assisted with the grammar, spelling and organization of this original work.

\balance
\printbibliography
\end{document}